\documentclass[12pt,a4paper]{article}
\usepackage{graphicx}
\pdfoutput=1

\title{\bf The high-energy radiation pattern from {\tt BFKLex} with double-log collinear contributions}
\author{G. Chachamis, A. Sabio Vera\\ \\
{\small Instituto de F{\' \i}sica Te{\' o}rica UAM/CSIC, Nicol{\'a}s Cabrera 15}\\ 
{\small \& Universidad Aut{\' o}noma de Madrid, E-28049 Madrid, Spain.}
}

\begin{document} 

\maketitle 

\abstract

We study high-energy jet production in the multi-Regge limit making use of the Monte Carlo event generator {\tt BFKLex} which includes collinear improvements in the form of double-log contributions as presented in~\cite{Chachamis:2015zzp}. Making use of the anti-$k_t$ jet algorithm in the {\tt FastJet} implementation, we present results for the average transverse momentum and azimuthal angle of the produced jets when two tagged forward/backward jets are present in the final state. We also introduce a new observable which accounts for the average rapidity separation among subsequent emissions. Results are presented, for comparison, at leading order and next-to-leading order, with the resummation of collinear double logs proposed in~\cite{Vera:2005jt}.

\section{Introduction}

The influence of the Balitsky-Fadin-Kuraev-Lipatov (BFKL) approach in final-state multi-jet configurations at the Large Hadron Collider (LHC) is a subject of intense debate in recent years. This approach applies when the center-of-mass energy is, in principle, asymptotically large ($\sqrt{s} \to \infty$). It can, however, affect hadron phenomenology pre-asymptotically at current LHC energies. One of the targets of the work here presented is to show how pre-asymptotic effects have sizable effects at present energies. This is the case for the simplest leading order (LO)~\cite{Lipatov:1985uk,Balitsky:1978ic,Kuraev:1977fs,Kuraev:1976ge,Lipatov:1976zz,Fadin:1975cb} as well as for the more sophisticated next-to-leading order (NLO)~\cite{Fadin:1998py,Ciafaloni:1998gs} and higher-order calculations.

For our presentation we focus on events where two forward/backward jets with rapidities $y_a$ and $y_b$ can be clearly identified. If the difference $Y=y_a-y_b$ is large enough then terms of the form $\alpha_s^n Y^n$ are important order-by-order to get a good description of measured cross sections which, in a nutshell, can be written in the factorized form 
\begin{eqnarray}
\sigma (Q_1,Q_2,Y) = \int d^2 \vec{k}_A d^2 \vec{k}_B \, {\phi_A(Q_1,\vec{k}_a) \, 
\phi_B(Q_2,\vec{k}_b)} \, {f (\vec{k}_a,\vec{k}_b,Y)}.
\end{eqnarray}
In this expression $\phi_{A,B}$ are impact factors depending on external scales, $Q_{1,2}$, and the off-shell reggeized gluon momenta, $\vec{k}_{a,b}$.  The gluon Green function $f$ depends on $\vec{k}_{a,b}$ and the center-of-mass energy in the scattering $\sim e^{Y/2}$.

For LHC phenomenology it is mandatory to work within the NLO approximation which introduces the dependence on physical scales such as the one associated to the running of the coupling and the one related to the choice of energy scale in the resummed logarithms~\cite{Forshaw:2000hv,Chachamis:2004ab,Forshaw:1999xm,Schmidt:1999mz}. It is possible to write the gluon Green function in an iterative way in transverse momentum and rapidity space at LO~\cite{Schmidt:1996fg} and NLO~ \cite{Andersen:2003an,Andersen:2003wy} in the form
\begin{eqnarray}
f &=& e^{\omega \left(\vec{k}_A\right) Y}  \Bigg\{\delta^{(2)} \left(\vec{k}_A-\vec{k}_B\right) + \sum_{n=1}^\infty \prod_{i=1}^n \frac{\alpha_s N_c}{\pi}  \int d^2 \vec{k}_i  
\frac{\theta\left(k_i^2-\lambda^2\right)}{\pi k_i^2} \nonumber\\
&&\hspace{-1.2cm}\int_0^{y_{i-1}} \hspace{-.3cm}d y_i e^{\left(\omega \left(\vec{k}_A+\sum_{l=1}^i \vec{k}_l\right) -\omega \left(\vec{k}_A+\sum_{l=1}^{i-1} \vec{k}_l\right)\right) y_i} \delta^{(2)} \hspace{-.16cm}
\left(\vec{k}_A+ \sum_{l=1}^n \vec{k}_l - \vec{k}_B\right) \hspace{-.2cm}\Bigg\}, 
 \end{eqnarray}
where 
\begin{eqnarray}
\omega \left(\vec{q}\right) &=& - \frac{\alpha_s N_c}{\pi} \log{\frac{q^2}{\lambda^2}} 
\end{eqnarray}
corresponds to the gluon Regge trajectory which carries a regulator, $\lambda$, of infrared divergencies. We have implemented this expression in the Monte Carlo event generator {\tt BFKLex} which we have already used for different applications ranging from collider phenomenology to more formal studies in the calculation of scattering amplitudes in supersymmetric theories~\cite{Chachamis:2013rca,Caporale:2013bva,Chachamis:2012qw,Chachamis:2012fk,Chachamis:2011nz,Chachamis:2011rw}. 

It turns out that the BFKL formalism can  be quite sensitive to collinear regions of phase space, in particular when the process-dependent impact factors are broad and allow for the external scales $Q_i$ to significantly deviate from the internal reggeized gluon transverse momenta $k_i$. In this case there exists a dominant double-log term in the NLO BFKL kernel in the collinear regions which takes the form
\begin{eqnarray}
\theta \left(k_i^2-\lambda^2\right) \to \theta \left(k_i^2-\lambda^2\right)  
-{\frac{\bar{\alpha}_s}{4} \ln^2{\left(\frac{\vec{k}_A^2}{\left(\vec{k}_A+\vec{k}_i\right)^2}\right)}},
\end{eqnarray}
which needs to be resummed to all orders to stabilize the behavior of the BFKL cross sections and to apply the formalism beyond the original multi-Regge kinematics. These issues have been investigated in~\cite{Salam:1998tj,Ciafaloni:2003ek}. In particular,  in~\cite{Vera:2005jt}, it was shown that the collinear corrections can be resummed to all-orders using the prescription
\begin{eqnarray}
\theta \left(k_i^2-\lambda^2\right) \to \theta \left(k_i^2-\lambda^2\right)  + \sum_{n=1}^\infty 
\frac{\left(-\bar{\alpha}_s\right)^n}{2^n n! (n+1)!} \ln^{2n}{\left(\frac{\vec{k}_A^2}{\left(\vec{k}_A+\vec{k}_i\right)^2}\right)}. 
\label{SumBessel}
\end{eqnarray}
As it was shown in~\cite{Vera:2005jt} this expression resums to a Bessel function of the first kind (similar results have recently been obtained in coordinate representation in~\cite{Iancu:2015vea}). Phenomenological applications of this resummation, not using a Monte Carlo approach, show agreement with experimental results and good perturbative convergence~\cite{Vera:2006un,Vera:2007kn,Caporale:2007vs,Vera:2007dr,Caporale:2008fj,Hentschinski:2012kr,Hentschinski:2013id,Caporale:2013uva,Chachamis:2015ona}.

In a recent work~\cite{Chachamis:2015zzp} we have shown how to implement this collinear resummation in  the {\tt BFKLex} Monte Carlo event generator and investigated what is its effect in the behavior of the gluon Green function. It is the target of this letter to extend that discussion to investigate the structure of the final state radiation, or, in other words, to which extend this collinear resummation affects the exclusive production of jets as obtained from the BFKL approach. We will show our results in the next Section, with a particular focus on how pre-asymptotic effects can be rather important in phenomenological applications. 

\section{Averages of characteristic quantities in multi-jet events}

This Section is devoted to the presentation of some numerical results as obtained from our Monte Carlo implementation of the NLO BFKL equation in the code {\tt BFKLex}. In particular, we investigate configurations with at least two jets, with transverse momentum $k_a$ and $k_b$,  in the final state, one very forward and the other very backward such that the rapidity spread between the two jets,
$y_a-y_b$, is large. This set up for the studied events is analogous to the Mueller-Navelet~\cite{Mueller:1986ey} configurations for which we would like in addition to probe the internal dynamics of the exchanged BFKL ladder and its associated mini-jet radiation. The advantage of this study is that the production of the two forward/backward tagged jets can be described via collinear factorization, thus reducing the uncertainty in the calculation of cross sections. This is relevant even though in the  present  letter we focus on presenting results at parton level. 

For the shake of definiteness, we introduce three quantities which are directly related to the jet activity along the ladder and uniquely (but not fully) characterize each event. These are three distinct averages for the jets in each event: of the modulus of their transverse momentum ($\langle p_t \rangle$), of their azimuthal angle ($\langle \theta \rangle$) and, finally, of the rapidity ratio ($\langle {\mathcal R}_y \rangle$) between subsequent jets. In more detail, let us assume an event for which, besides the two tagged jets with transverse momentum $k_a$ and $k_b$, there exists a number $N$ of further final-state jets. For each of these jets (whether and when we can consider them as mini-jets will be discussed in the following) we can define three variables: the modulus of its transverse momentum, $|k_i|$,
its azimuthal angle $\theta_i$ and its rapidity $y_i$, with $1 \le i \le N$. Then, the average transverse momentum, azimuthal angle and rapidity ratio would, respectively, read:
\begin{eqnarray}
\langle p_t \rangle &=& \frac{1}{N} \sum_{i=1}^{N} |k_i|;
\label{eq:observable1}\\
\langle \theta \rangle &=& \frac{1}{N} \sum_{i=1}^{N} \theta_i;
\label{eq:observable2}\\
\langle {\mathcal R}_y \rangle &=& \frac{1}{N+1}  \sum_{i=1}^{N+1} \frac{y_i}{y_{i-1}}.
\label{eq:observable3}
\end{eqnarray}
In these definitions we set $y_0 = y_a$, $y_{N+1} = y_b = 0$ and $y_{i-1} > y_i$.
All three observables are tailored such that should give an accurate view 
of how closely we follow the multi-Regge kinematics in our Monte Carlo solution
of the BFKL equation. We should keep in mind that the
multi-Regge kinematics dictates similar transverse sizes for all jets, 
a strong ordering in rapidities, $y_{i-1} \gg y_i$ and generally, 
azimuthal angles with equal probability in the range $[0, 2 \pi)$.

In order to have a better control on how these observables behave in a collider experimental setup, 
we use the anti-$k_t$ jet algorithm~\cite{Cacciari:2008gp}
in the {\tt FastJet} implementation~\cite{Cacciari:2011ma,hep-ph/0512210}
with a jet radius of $R=0.7$ for the emitted jets in the final state. For the presentation of our numerical results, we have considered two different configurations for the transverse momenta
of the forward/backward jets:
i) $k_a = 10$ GeV, $k_b = 12$ GeV,
ii) $k_a = 10$ GeV, $k_b = 20$ GeV and three different rapidity
differences $y_a-y_b = 4, 6, 8$. For each of these cases we have run {\tt BFKLex} and produced differential distributions for the observables in Eqs.~(\ref{eq:observable1}),~(\ref{eq:observable2}),
and~(\ref{eq:observable3}).
Two of the observables here discussed have boundary values $0 \le \langle \theta_t \rangle < 2 \pi$ and
$0 \le \langle {\mathcal R}_y \rangle \le 1$.
We sliced the allowed $\langle \theta \rangle$ and $\langle {\mathcal R}_y \rangle$
ranges into 100 bins and the $\langle p_t \rangle$ range into 300 bins.
Any produced event by {\tt BFKLex} characterised by $\langle p_t \rangle_{bin[m]}$,
$\langle \theta \rangle_{bin[n]}$ and $\langle {\mathcal R}_y \rangle_{bin[l]}$ adds
its weight to the $m$-th bin of the $\langle p_t \rangle$ distribution, the 
$n$-th bin of the $\langle \theta_t \rangle$ distribution and the 
$l$-th bin of the $\langle  {\mathcal R}_y  \rangle$ distribution such that the area
under any differential distribution gives the full gluon Green function.

\begin{figure}
\begin{center}
\includegraphics[height=6cm]{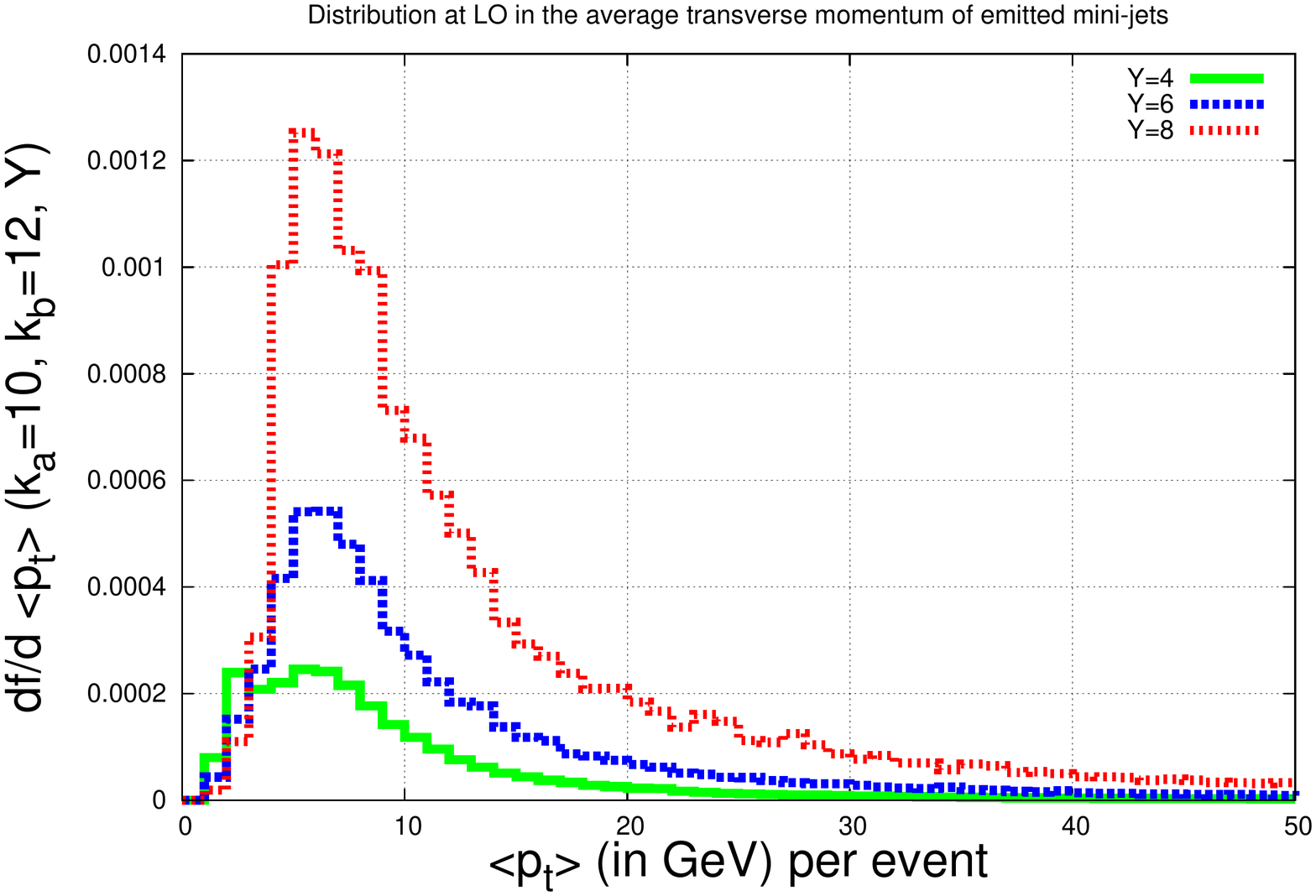}\includegraphics[height=6cm]{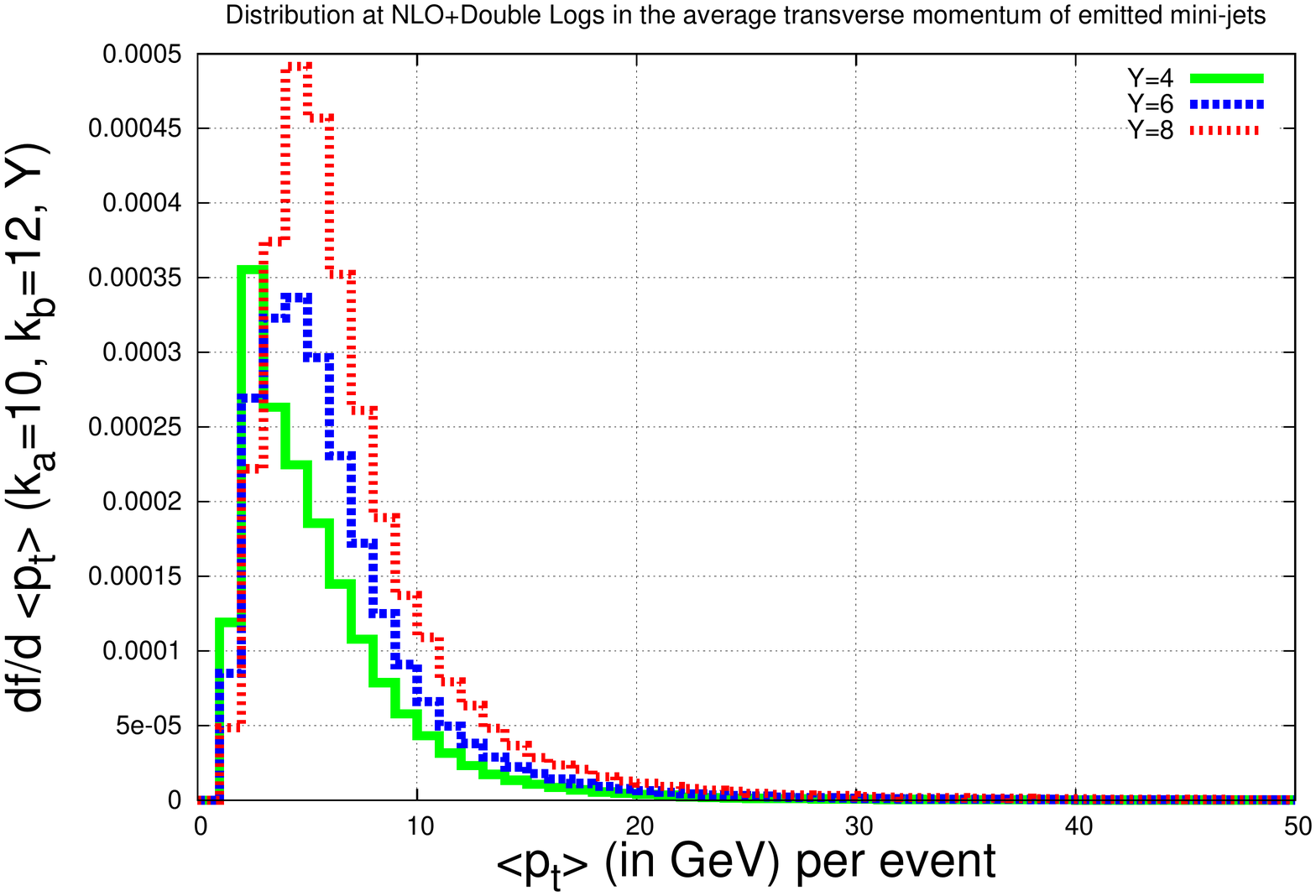}\\
\includegraphics[height=6cm]{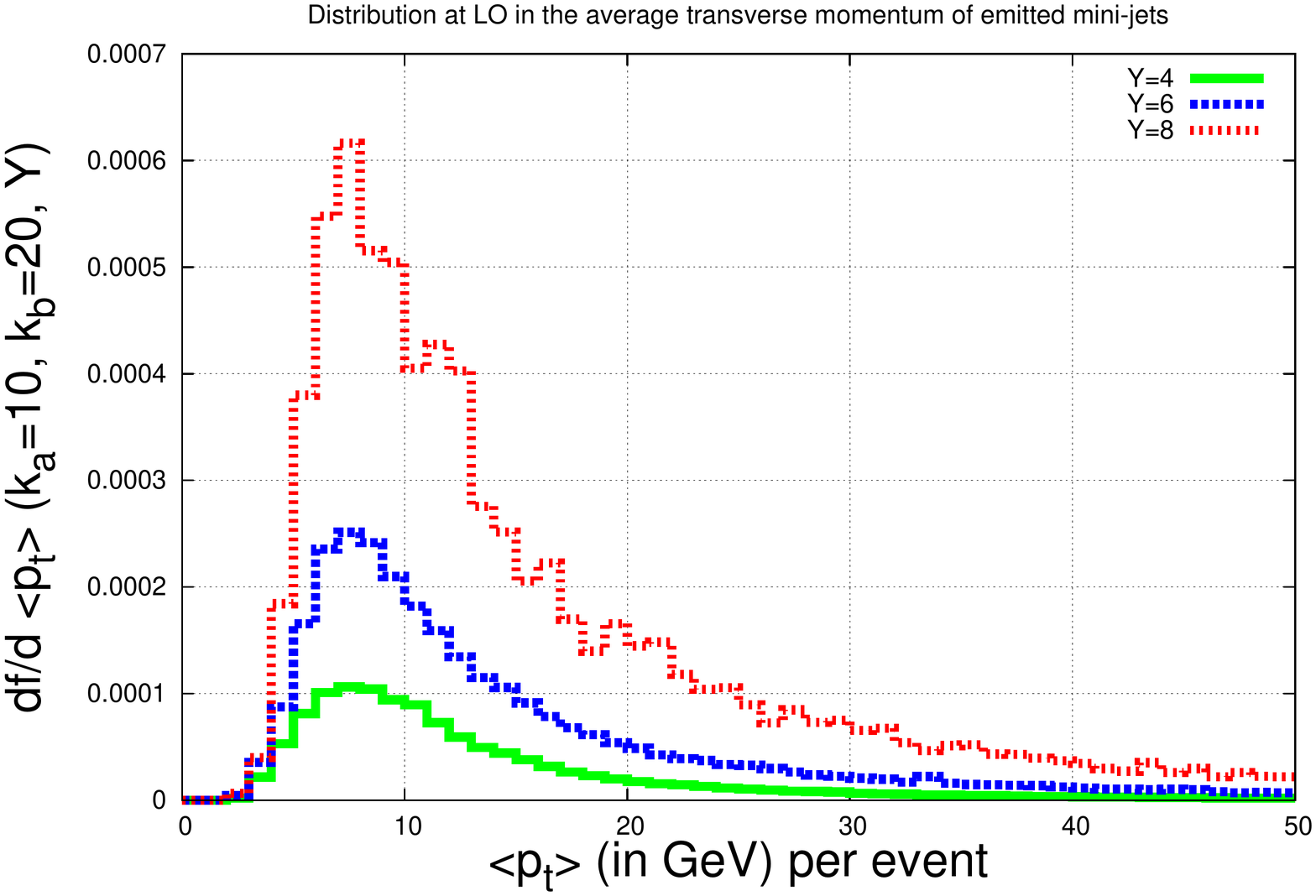}\includegraphics[height=6cm]{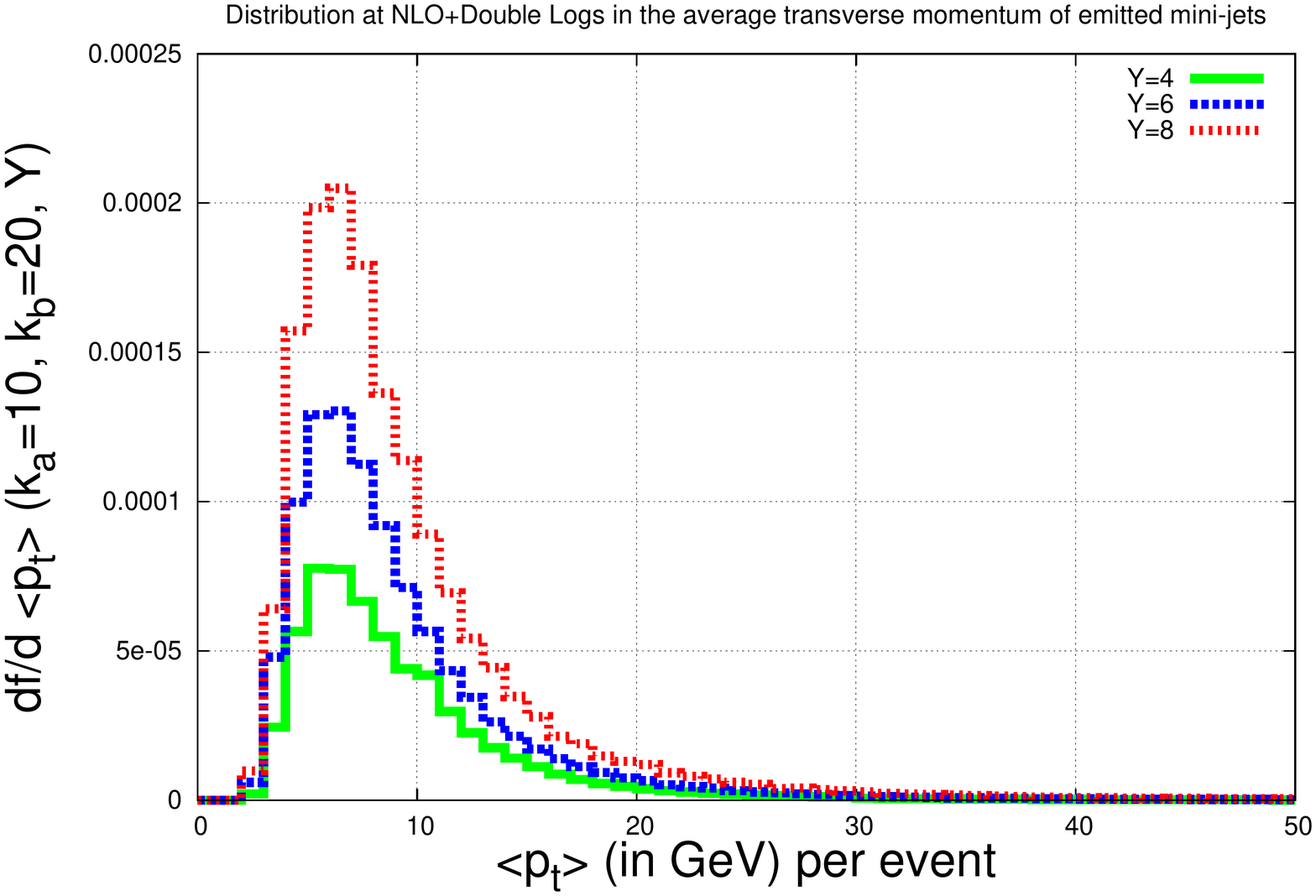}\\
\includegraphics[height=6cm]{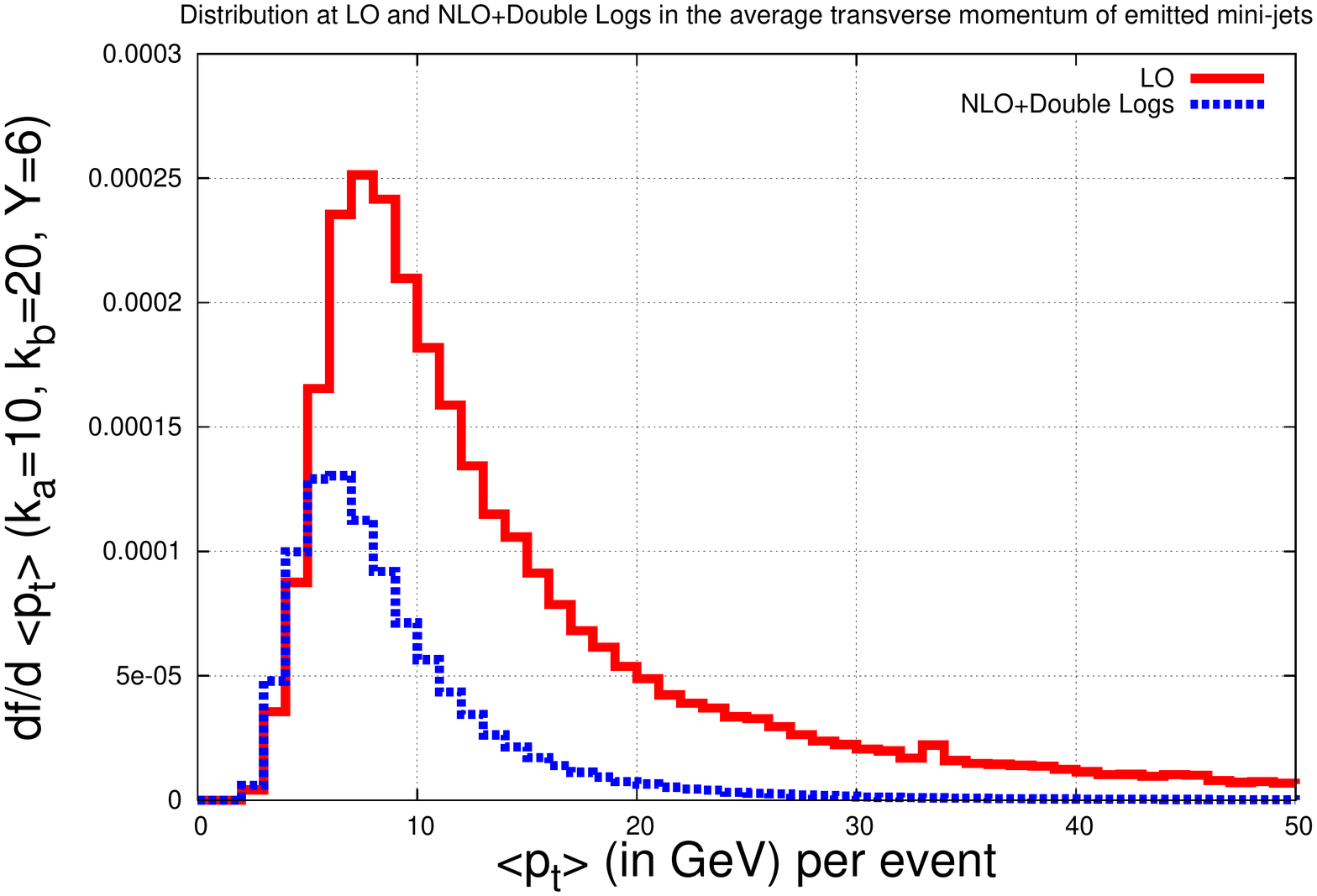}
\end{center}
\vspace{-1cm}
\caption{Distribution at LO (top/middle left) and NLO+Double Logs (top/middle right) in the average transverse momentum of emitted mini-jets per event, for different values of the transverse momentum of the most forward/backward tagged jets. In the plot at the bottom we compare the LO to the NLO+Double Log distribution for $Y=6$.}
\label{PtPlots}
\end{figure}

We have gathered together our results at LO and NLO+Double Logs in three figures. 
In  Fig.~\ref{PtPlots}. 
we show the differential distribution at LO (top/middle left) and NLO+Double Logs (top/middle right) in the average transverse momentum $\langle p_t \rangle$ of emitted mini-jets per event, for different values of the transverse momentum of the most forward/backward tagged jets. Having as a common value $k_a=10$ what we find are broad distributions with a maximal value at $\langle p_t \rangle \simeq 6$ GeV for $k_b=12$ and 8 GeV for $k_b=20$. These maxima are independent of the value of the total rapidity span in the process when we evaluate the process at LO and when including the higher order corrections. This is one of the kinematical conditions defining MRK: to have a non-growing with $s$ invariant mass for the produced jets. 

 As we can see in the bottom plot of Fig.~\ref{PtPlots} the average $p_t$ is smaller when considering NLO contributions together with double logarithmic collinear terms than when we operate at LO. It is noteworthy to comment on the broadness of the distributions. A large contribution to the Green function and, hence, to cross sections stems from jets produced with a large transverse momentum which cannot be considered any longer as mini-jets. This feature is clearly seen at LO and gets reduced in higher-order calculations. This reduction is related of the shrinkage of the diffusion picture at LO+Double Logs shown in our previous work in~\cite{Chachamis:2015zzp}. The areas under the different distributions are much larger at LO for any $Y$ due to the strong suppression of the Pomeron intercept when going beyond LO. 

\begin{figure}
\begin{center}
\includegraphics[height=6cm]{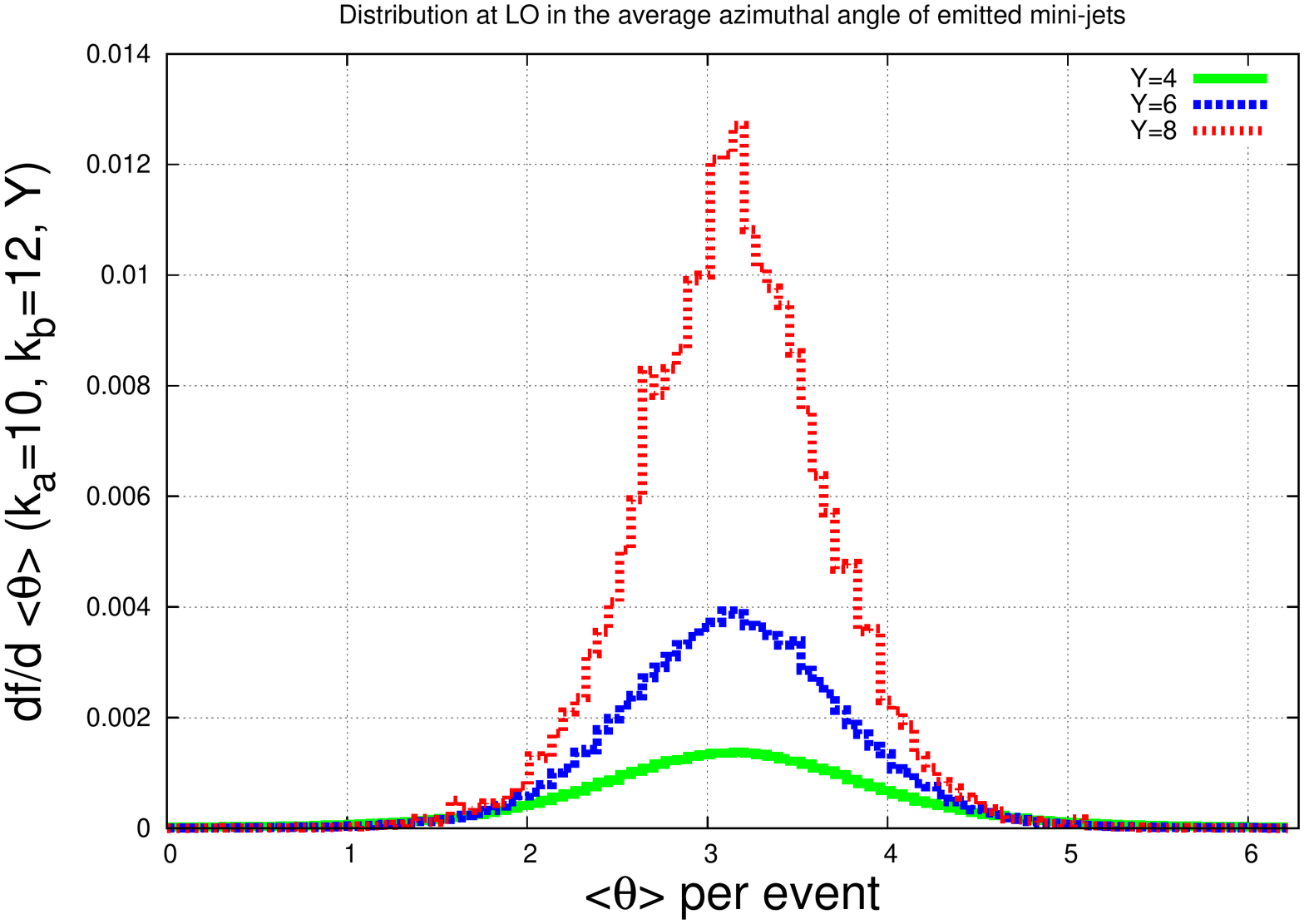}\includegraphics[height=6cm]{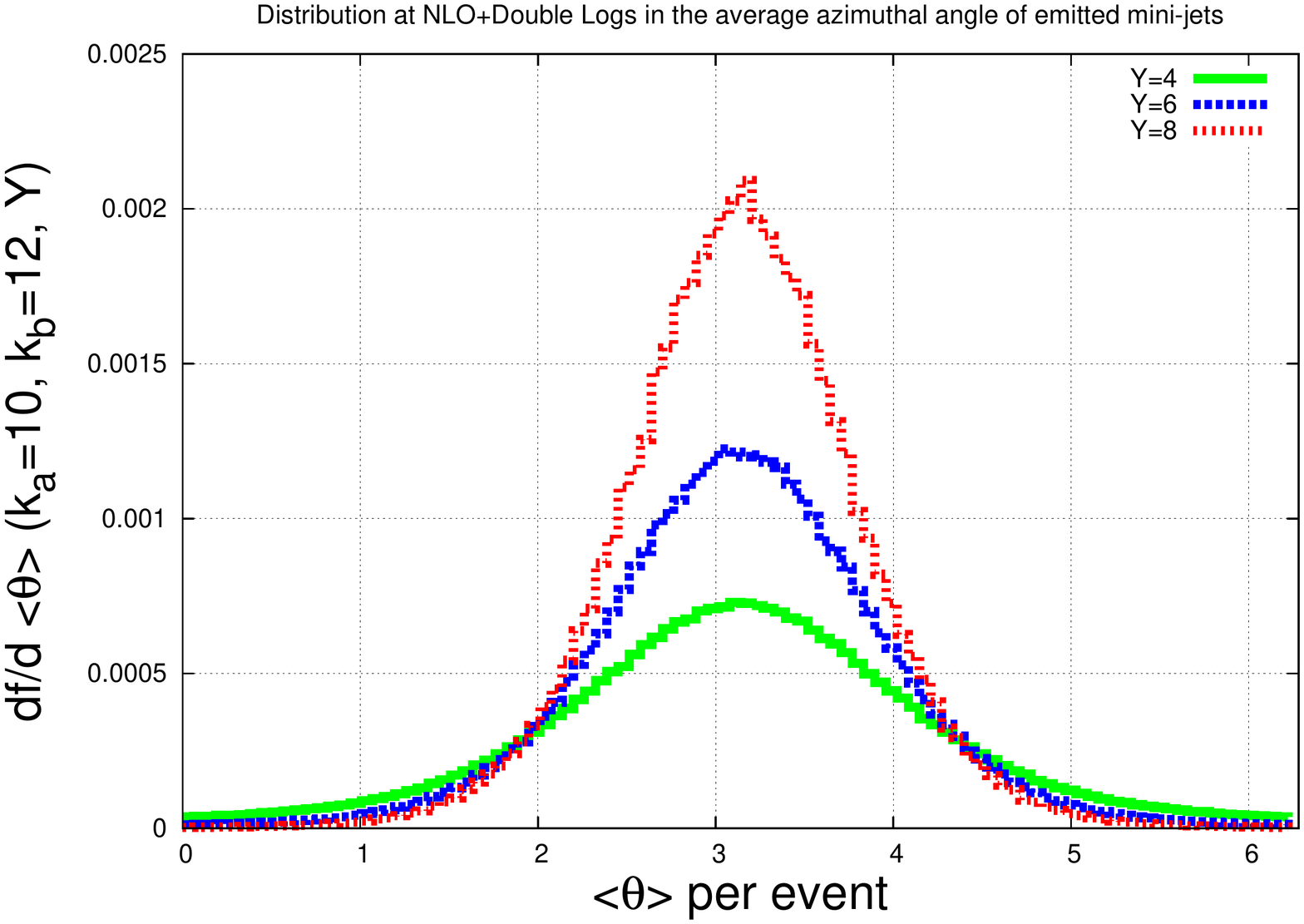}\\
\includegraphics[height=6cm]{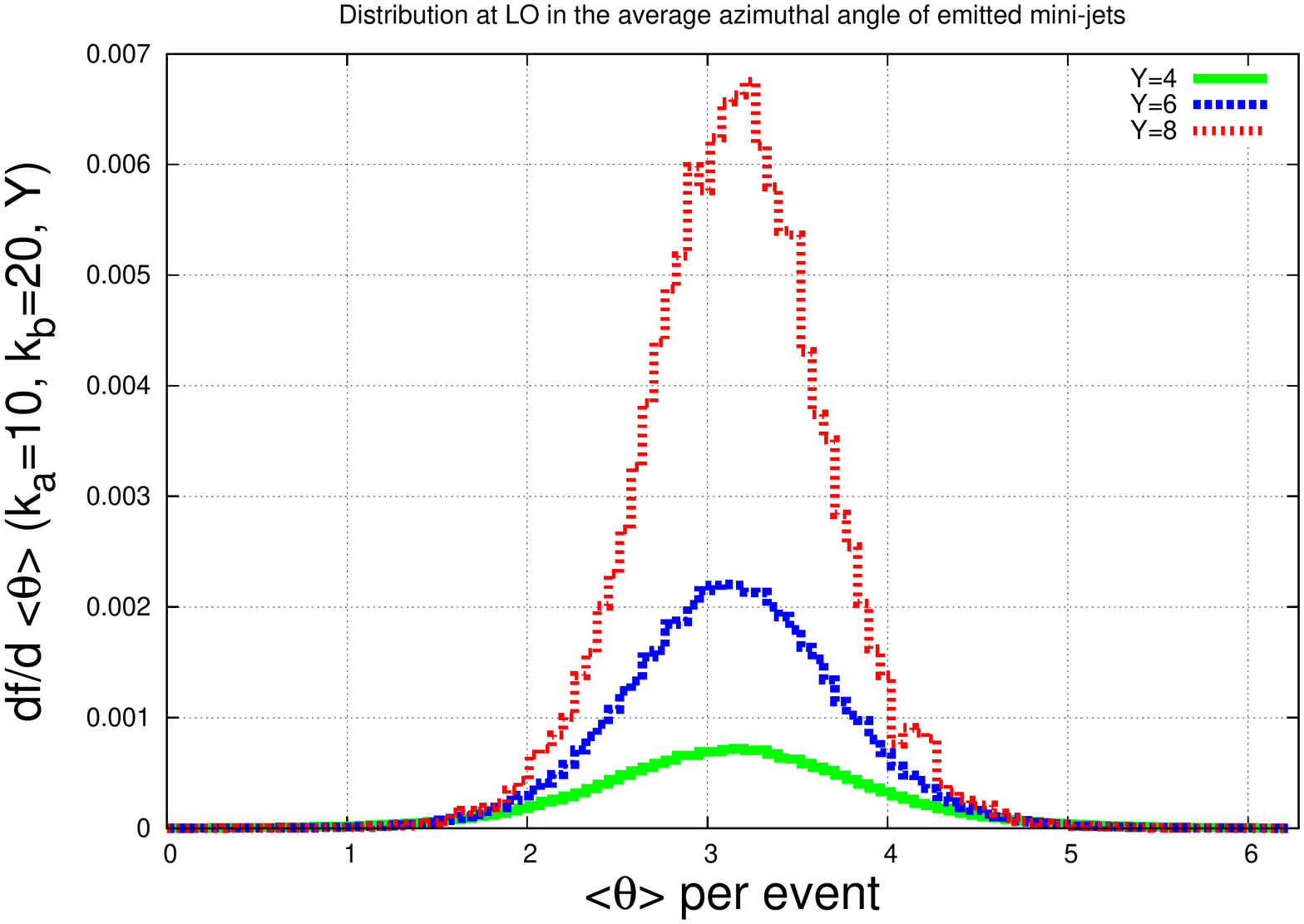}\includegraphics[height=6cm]{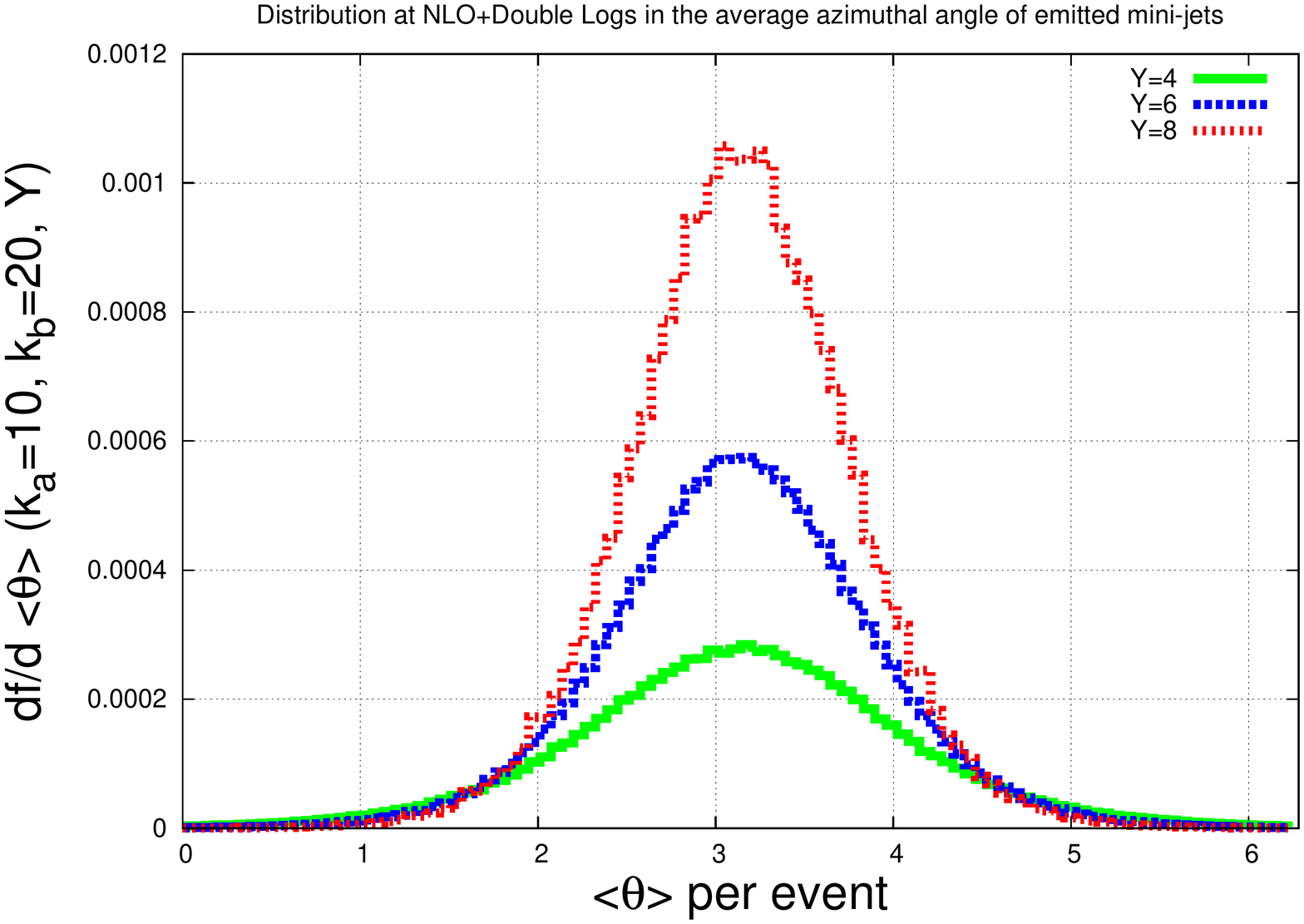}\\
\includegraphics[height=6cm]{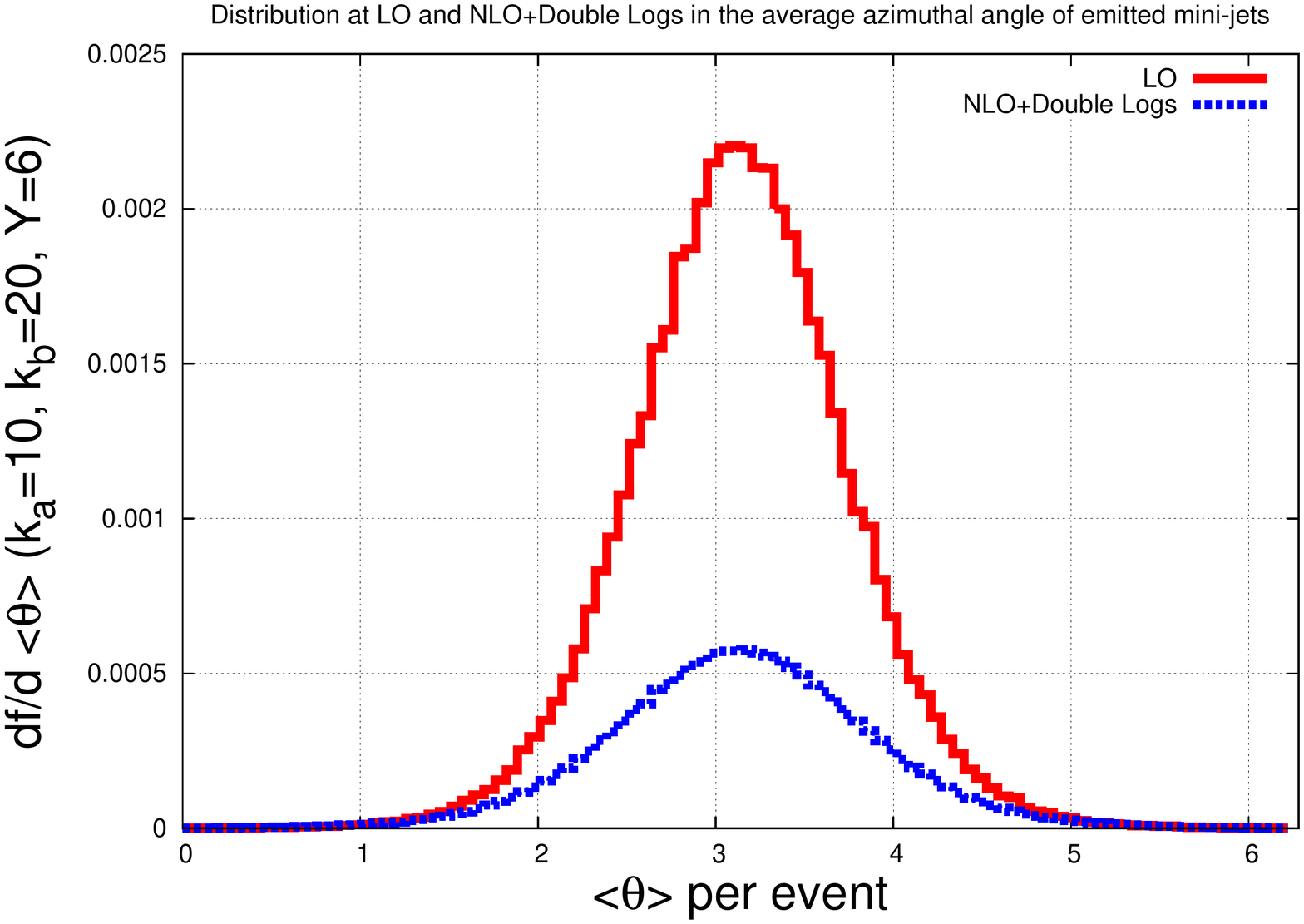}
\end{center}
\vspace{-1cm}
\caption{Distribution at LO (top/middle left) and NLO+Double Logs (top/middle right) in the average azimuthal angle of emitted mini-jets per event, for different values of the transverse momentum of the most forward/backward tagged jets. In the plot at the bottom we compare the LO to the NLO+Double Log distribution for $Y=6$.}
\label{thetaPlots}
\end{figure}

Note that to produce the distributions in this work we have taken a random value for the azimuthal angle between the two most forward/backward tagged jets which changes event by event. We can then investigate the average angle $\langle \theta \rangle$ per event at which the remaining jets will be produced. This is shown in Fig.~\ref{thetaPlots}.
At LO the bulk of the radiation carries an average angle in between $ \simeq \pi \pm 1$ which does not vary when $Y$ changes. This is also true when higher-order corrections are included in the analysis. 

\begin{figure}
\begin{center}
\includegraphics[height=6cm]{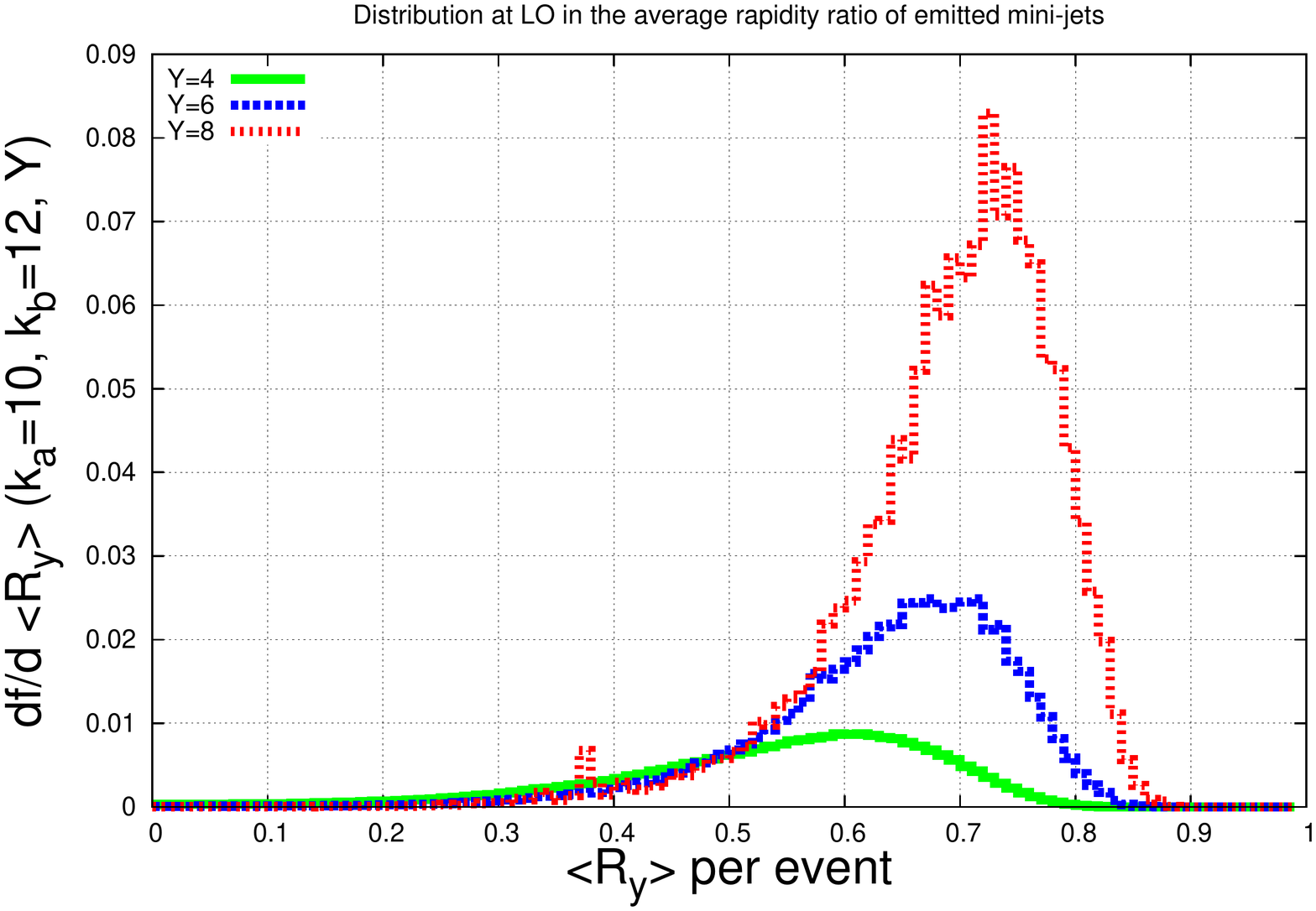}\includegraphics[height=6cm]{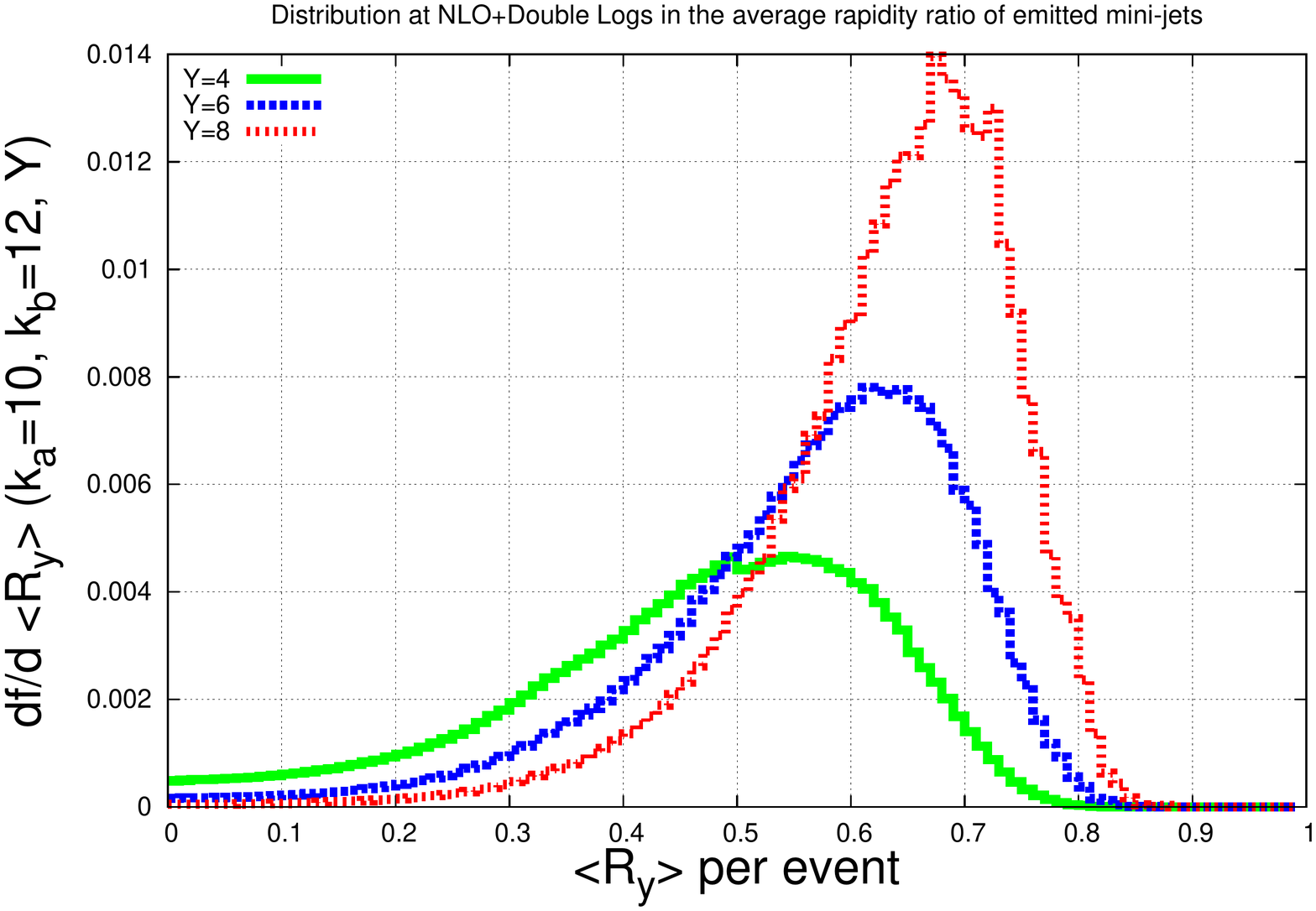}\\
\includegraphics[height=6cm]{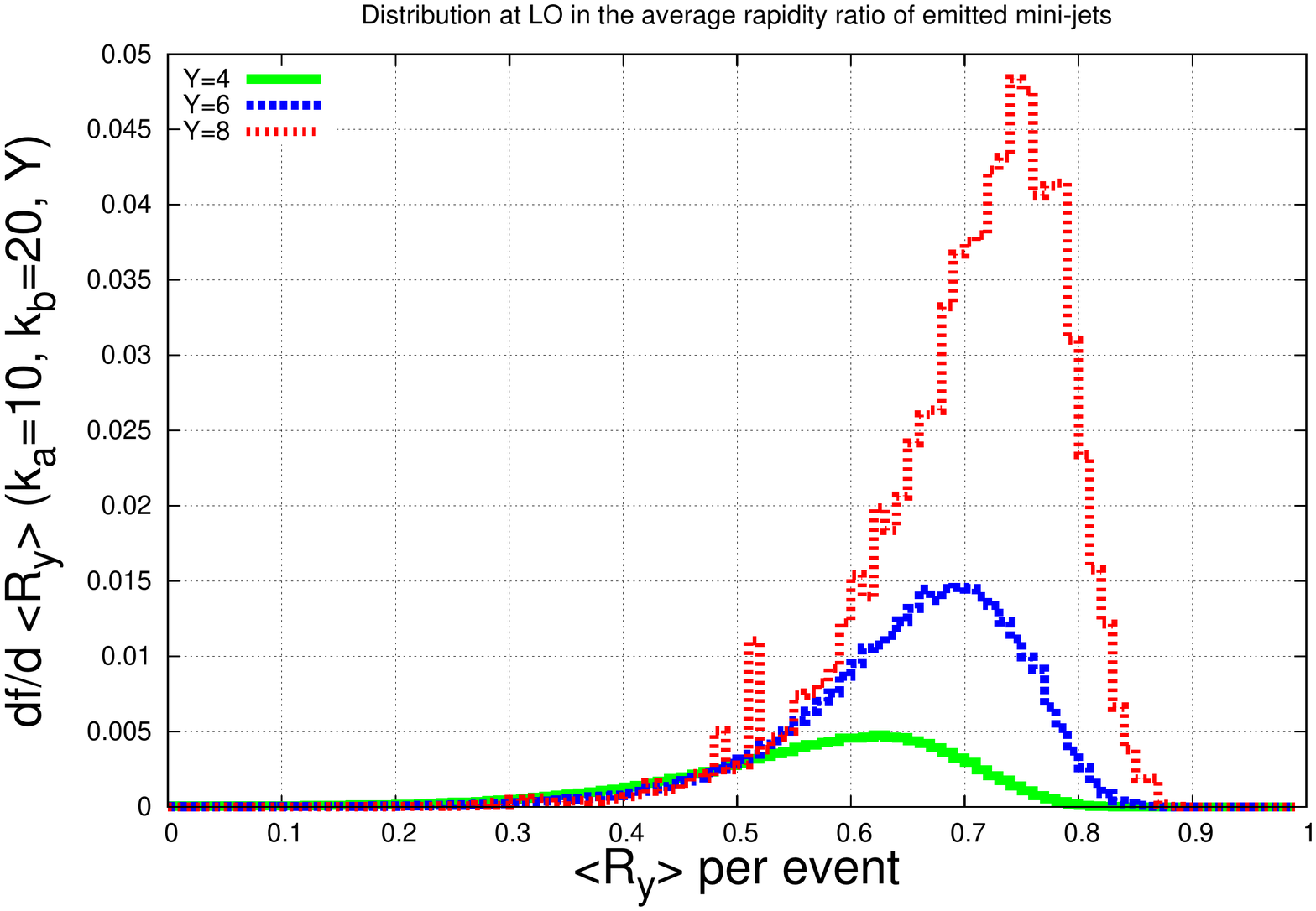}\includegraphics[height=6cm]{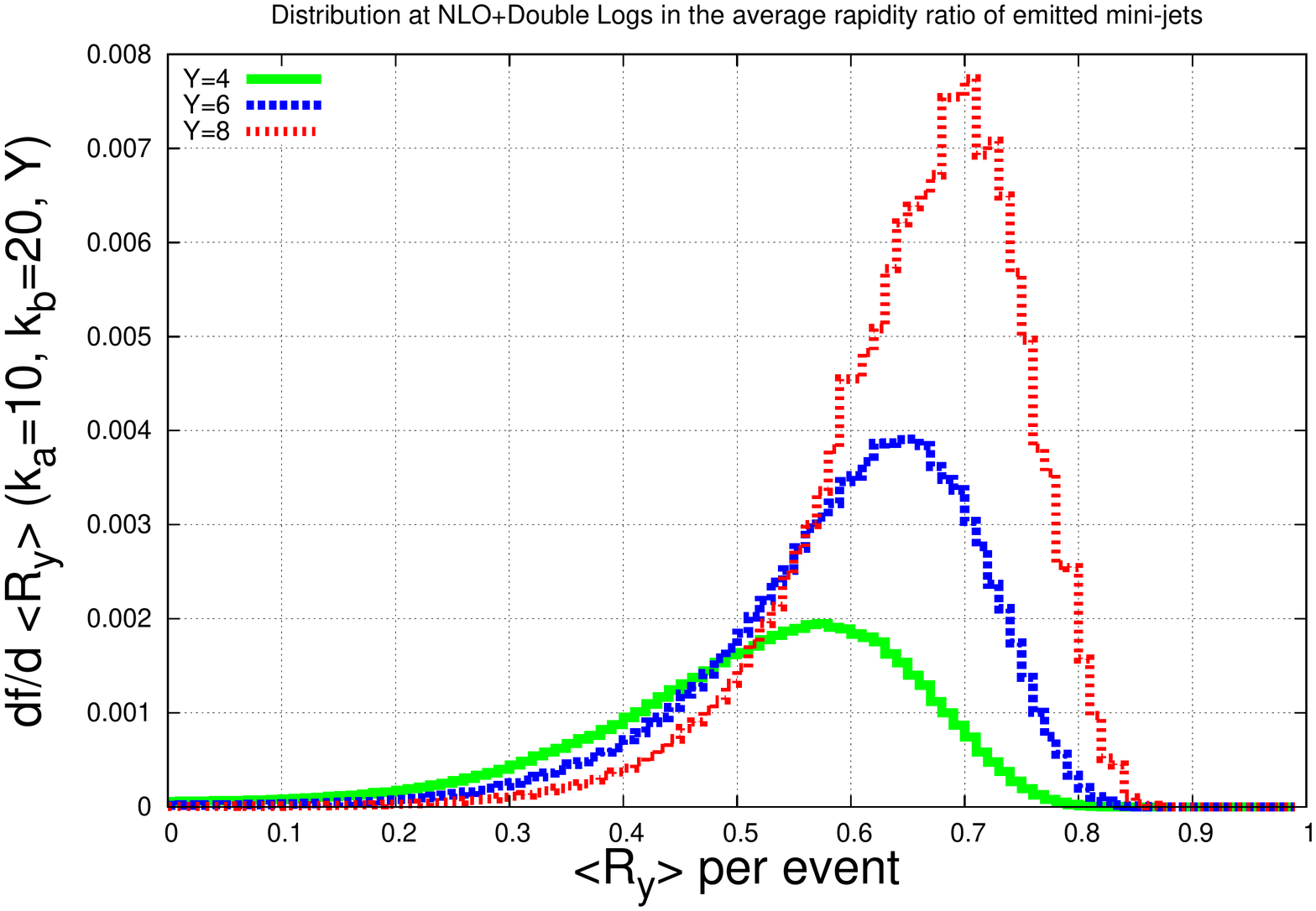}\\
\includegraphics[height=6cm]{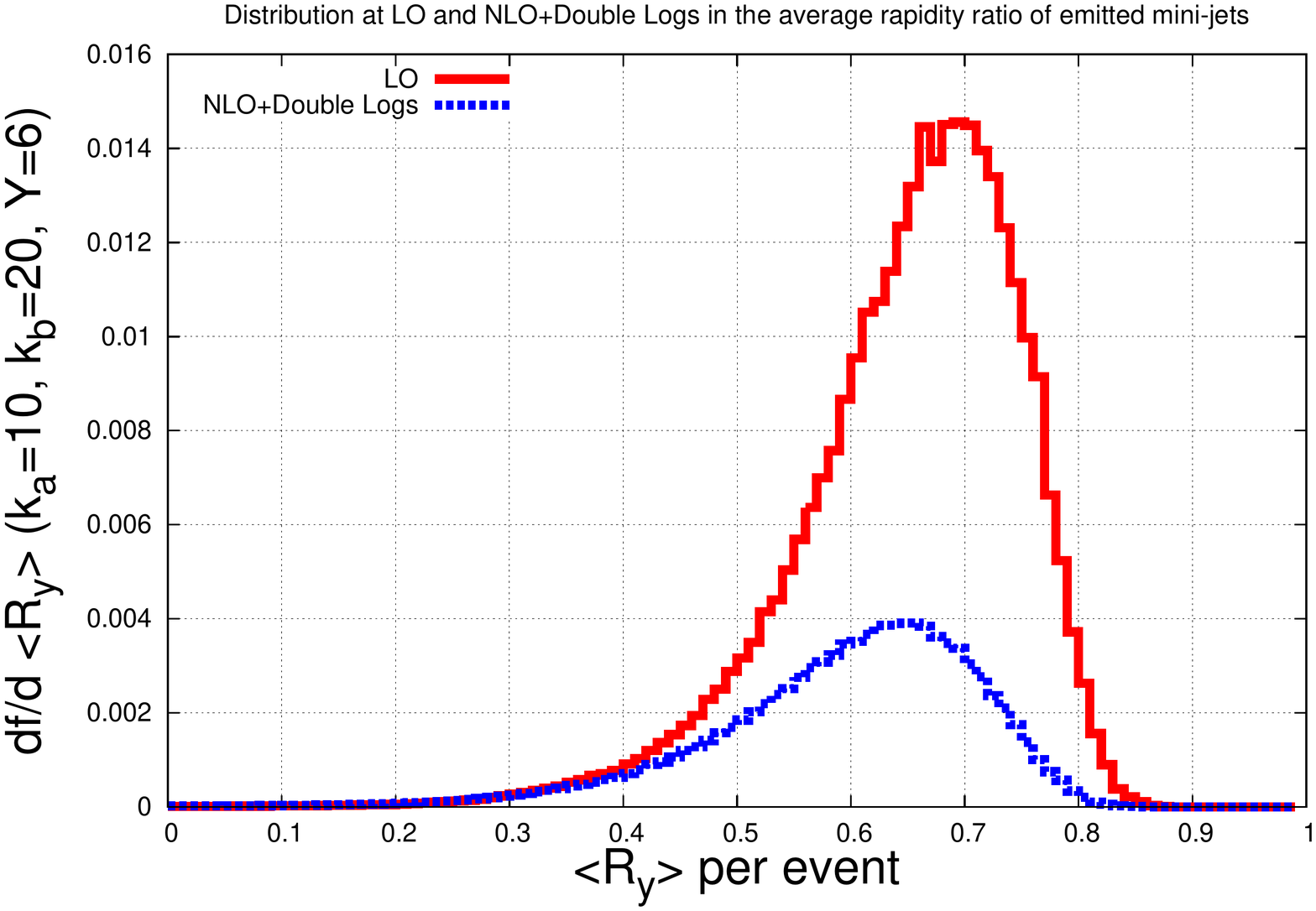}
\end{center}
\vspace{-1cm}
\caption{Distribution at LO (top/middle left) and NLO+Double Logs (top/middle right) in rapidity ratios of emitted mini-jets per event, for different values of the transverse momentum of the most forward/backward tagged jets. In the plot at the bottom we compare the LO to the NLO+Double Log distribution for $Y=6$.}
\label{rapPlots}
\end{figure}

Let us now investigate the mean distance in rapidity between emissions in the BFKL ladder. We have addressed this point in Fig.~\ref{rapPlots} where the ratio $\langle {\mathcal R}_y \rangle$ has been numerically investigated in detail.  
We observe that the differential distributions for these ratios have their maximal contribution for $\langle {\mathcal R}_y \rangle$ larger than 0.5. Since these distributions are broad, this implies that there are substantial contributions to the cross section from kinematical (preasymptotic) configurations away from MRK. In an asymptotic MRK typical event we can consider the rapidity span to be populated by $N$ emissions equally spaced at rapidity intervals of length $\Delta$ in such a way that $Y = (N+1) \Delta$ (it is well-known that at asymptotically large energies $\Delta = 1/{\bar{\alpha}_s \omega_0}$ with $\omega= 4 \ln 2$, the LO Pomeron intercept. In our case with $\bar{\alpha}_s = 0.2$ this implies asymptotic $\Delta^{\rm as}_{\rm LO} \simeq 1.8$ and $\Delta^{\rm as}_{\rm NLO+Double Logs} \simeq 3.3$). Hence, the rapidity corresponding to the $i$-th emission is $(N+1-i) \Delta$ and our average ratio becomes independent of $\Delta$ reading
\begin{eqnarray}
\langle {\mathcal R}_y \rangle_{\rm MRK} &=& \frac{1}{N+1} \sum_{i=1}^{N} \frac{i}{i+1} ~=~
\frac{N+1- \psi (N+2) + \psi (1)}{N+1} \nonumber\\
&=& 1+ \frac{\Delta}{Y } \Bigg(\psi (1) -\psi \left(1+\frac{Y}{\Delta}\right) \Bigg)\nonumber\\
&\simeq& 1 + \frac{\Delta}{Y} \left( \psi (1) +\ln{\frac{\Delta}{Y}}\right) + \cdots 
\end{eqnarray}
The last approximation is valid when $\Delta \ll Y$. At LO we need a $\Delta_{\rm LO} \simeq 0.6196$ to reproduce all the maxima of the distributions for different $Y$ ($\langle {\mathcal R}_y \rangle_{\rm MRK} \simeq 0.61, 0.70, 0.75$ at $Y=4,6,8$, respectively). At NLO this number increases to $\Delta_{\rm NLO+Double Logs} \simeq 0.8097$ ($\langle {\mathcal R}_y \rangle_{\rm MRK} \simeq 0.54, 0.64, 0.70$ at $Y=4,6,8$, respectively) (these numbers are quite similar for the range of momenta here discussed) . This is consistent with having a reduced multiplicity in the final state when the higher-order corrections are taken into account. As we increase the available scattering energy these MRK configurations become more relevant when constructing the gluon Green function but at phenomenological rapidity differences we can see from our analysis that other kinematical regions also play an important role. An example of a kinematical set up with a smaller / larger $\langle {\mathcal R}_y \rangle$  than its maximal value could be a final state with all emissions at equidistant rapidities apart from one pair of jets whose relative ratio of rapidities could be particularly smaller (close to 0) / larger (close to 1) than the others.

This concludes our discussion of some of our numerical results. We find that the observables here presented are worth of experimental investigation at the LHC. It is important to establish if the pre-asymptotic effects are already present in the data. The characteristic broadening of the distributions that we have shown for $\langle p_t \rangle$, $\langle \theta \rangle$ and $\langle {\mathcal R}_y \rangle$ is a distinct signal of BFKL activity and should be put forward for experimental verification. In particular, since the collinear emissions do play a role in the form of double-log contributions, it would be interesting to gauge their importance at LHC data. 

\section{Summary \& Outlook}

We have presented a set of observables characterizing multi-jet configurations event by event (average transverse momentum, average azimuthal angle, average ratio of jet rapidities) which can be used to find distinct signals of BFKL dynamics at the LHC. A numerical analysis has been shown using the Monte Carlo event generator {\tt BFKLex}, modified to include higher-order collinear corrections in addition to the transverse-momentum implementation of the NLO BFKL kernel and the anti-$k_t$ jet algorithm as in {\tt FastJet}. In order to have a cleaner theoretical background within collinear factorization we demand to always have two tagged forward/backward jets in the final state. 

The advantage of the LHC to study this type of physics is the large available center-of-mass energy together with high statistics, which allow for the possibility to apply strong kinematical cuts. These cuts are the key to pin down the multi-Regge kinematics and to propose new observables capable of discriminating this region of phase space from other, more conventional, ones. It is mandatory to make use of very exclusive observables in order to find a precise window of applicability of the BFKL formalism which can then safely be extended to other, less restrictive, experimental setups.

\begin{flushleft}
{\bf \large Acknowledgements}
\end{flushleft}
G.C. acknowledges support from the MICINN, Spain, under contract FPA2013-44773-P. 
A.S.V. acknowledges support from Spanish Government (MICINN (FPA2010-17747,FPA2012-32828)) and to the Spanish MINECO Centro de Excelencia Severo Ochoa Programme (SEV-2012-0249).

\end{document}